# A tale of clusters: No resolvable periodicity in the terrestrial impact cratering record

Matthias M. M. Meier[1]* and Sanna Holm-Alwmark[2]

[1]ETH Zurich, Institute of Geochemistry & Petrology, Clausiusstrasse 25, 8092 Zurich, Switzerland
[2]Department of Geology, Lund University, S olvegatan 12, 22362 Lund, Sweden

* E-Mail: matthias.meier@erdw.ethz.ch

**ABSTRACT**

Rampino & Caldeira (2015) carry out a circular spectral analysis (CSA) of the terrestrial impact cratering record over the past 260 million years (Ma), and suggest a ~26 Ma periodicity of impact events. For some of the impacts in that analysis, new accurate and high-precision ("robust"; 2SE<2%) $^{40}Ar$-$^{39}Ar$ ages have recently been published, resulting in significant age shifts. In a CSA of the updated impact age list, the periodicity is strongly reduced. In a CSA of a list containing only impacts with robust ages, we find no significant periodicity for the last 500 Ma. We show that if we relax the assumption of a fully periodic impact record, assuming it to be a mix of a periodic and a random component instead, we should have found a periodic component if it contributes more than ~80% of the impacts in the last 260 Ma. The difference between our CSA and the one by Rampino & Caldeira (2015) originates in a subset of "clustered" impacts (i.e., with overlapping ages). The ~26 Ma periodicity seemingly carried by these clusters alone is strongly significant if tested against a random distribution of ages, but this significance disappears if it is tested against a distribution containing (randomly-spaced) clusters. The presence of a few impact age clusters (e.g., from asteroid break-up events) in an otherwise random impact record can thus give rise to false periodicity peaks in a CSA. There is currently no evidence for periodicity in the impact record.

## 1 INTRODUCTION

Is there a periodic pattern in biosphere mass extinction events and the formation of large impact structures on Earth? This question has now been debated for more than three decades (e.g., Alvarez and Muller, 1984; Bailer-Jones, 2011, 2009; Grieve et al., 1985; Melott and Bambach, 2010). In a recent publication, Rampino and Caldeira, (2015) listed the ages of 37 terrestrial impact structures (hereafter: "impacts"), drawn from an online compilation known as the Earth

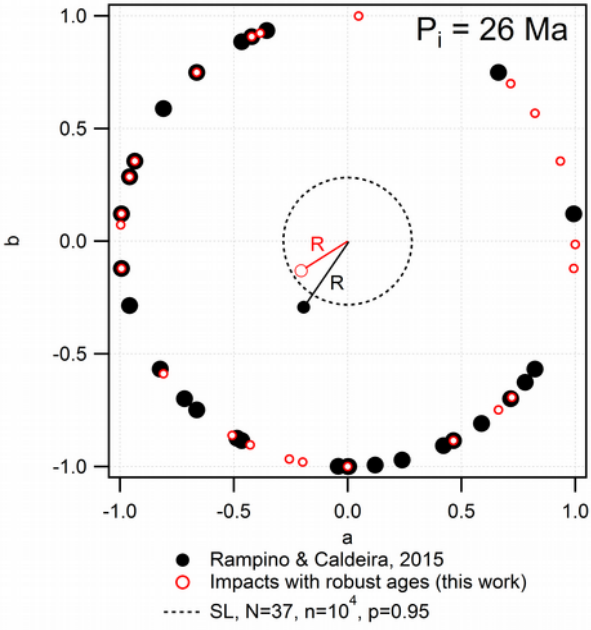

*Figure 1.* The time-line of impact ages as given in the list by Rampino & Caldeira (2015) (solid symbols) and in this work (all robust ages from Table 1; open symbols), wrapped around a circle with a circumference of 26 Ma. The symbols connected to the origin with a solid line represent the 2-dimensional arithmetic averages, the length of this line representing R. The short-dashed line represents the significance limit (SL).

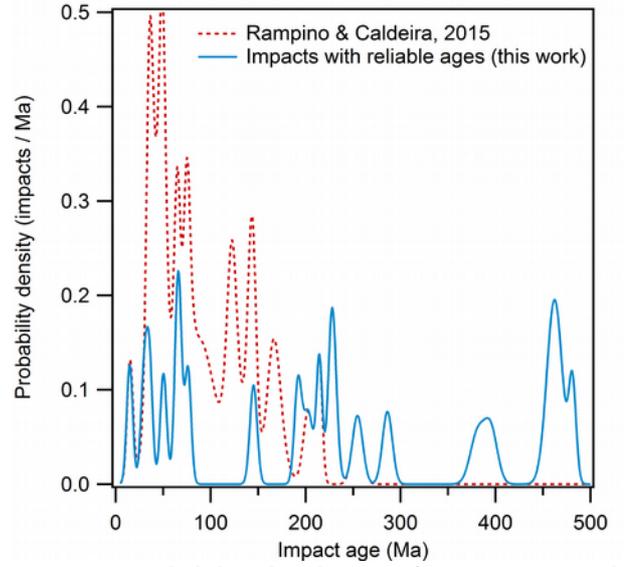

*Figure 2.* Probability distribution of impact ages in the list compiled by Rampino & Caldeira (2015) (short-dashed line) and in this work (solid line; Table 1). To guide the eye, both distributions have been smoothed by adding a Gaussian-distributed uncertainty with a standard deviation of 3 Ma.

Impact Database (EIDB). They listed all impacts in the 5 – 260 Ma age range with age uncertainties <±10 Ma, as given in the EIDB. They then carried out a so-called circular spectral analysis (CSA) to determine whether that record follows a periodic pattern, and found a significant peak at a period of 25.8±0.6 Ma. In a CSA, the time-line of impact ages is wrapped around a circle with a circumference corresponding to a trial period $P_i$ (see Fig. 1). For each trial period $P_i$, the value R is determined, which is given by the length of the vector connecting the origin with the data point representing the 2-dimensional arithmetic average of all data points in the series (neglecting their individual uncertainties). If the data points concentrate preferentially in one region of the circle, e.g., due to a strong periodicity of the record at $P_i$, the average of the series will plot close to this region and away from the origin, and thus R will be large. The angle between that region of concentration and the [0,1] direction then represents the phase of the periodic function (i.e., the time since the last maximum). On the other hand, if the data points are distributed randomly along the circumference of the circle at $P_i$, the average of the data points will plot close to the origin, and R will be small. Plotting R versus different values of $P_i$ yields a so-called periodogram (Lutz, 1985; Rampino and Caldeira, 2015), which we will also use here (with 5 Ma < $P_i$ < 50 Ma, and $P_{i+1}$–$P_i$ = 0.1 Ma). The value that R has to exceed (for a period/circumference $P_i$) before we can start suspecting a possible periodicity at $P_i$ is called the significance limit (SL; short-dashed line in Fig. 1). It has typically been set by the 95[th] percentile of R values (for each $P_i$) drawn from a large number (in Rampino and Caldeira (2015) and here: $10^4$) of artificial impact age series, containing the same number of impacts as the series which is tested for periodicity, but with impact ages following a random distribution. The CSA approach has seen widespread use, but it has also been criticised as being overly simplistic. As pointed out in detail by Bailer-Jones (2011), what is actually being tested by a CSA is the significance of a certain periodic model relative to a certain random model. Rejection of the random model does not necessarily mean that the periodic model is the best explanation for the observed impact record – instead, there might be other models which are more likely than both. In this work, we will use the CSA primarily as a didactic tool, to demonstrate that even if we accept the CSA as a valid approach, there is no periodicity in the current impact record if only robust (i.e., both accurate and precise) impact ages are used.

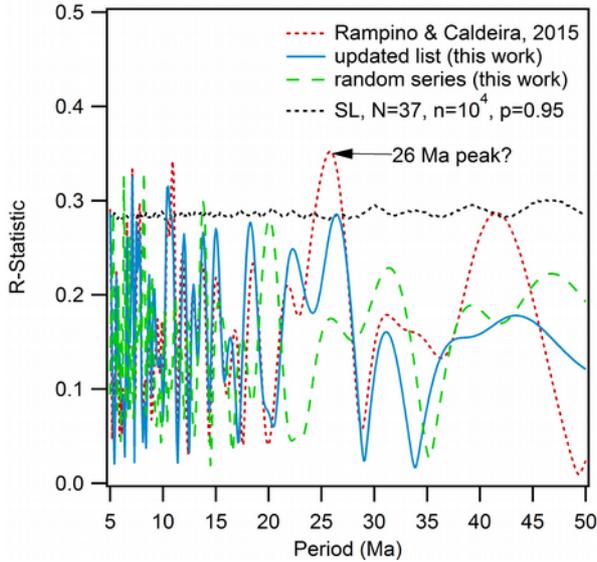

*Figure 3.* Circular Spectral Analysis (CSA) for the original list of impacts given by Rampino & Caldeira (2015) (short-dashed line), for the updated list (solid line), and for an artificial series of impact ages (long-dashed line).

The Earth Impact Database used by Rampino and Caldeira (2015) is a compilation of published and commonly cited ages maintained by the Planetary and Space Science Centre (PASSC) of the University of New Brunswick. As the data listed in the EIDB are derived from a large variety of sources and are therefore of variable quality, impact ages drawn from the EIDB require a careful case-by-case evaluation before they can be analysed as a population. The need for this is exemplified by three impacts in the list given by Rampino and Caldeira (2015): (1) the age of Carswell given in the EIDB (115±10 Ma) has recently been revised upwards to 485.5±1.5 Ma (Bleeker et al. 2015; Alwmark et al., submitted; note that all robust ages we will give here are based on the revised $^{40}$K decay constant by Renne et al. (2010); (2) the age of Puchezh-Katunki has recently been revised to 193.8±1.1 Ma, significantly higher than the commonly cited age of 167±3 (Holm-Alwmark et al., submitted); (3) the age of Logoisk has been revised to 30.0±0.5 Ma by Sherlock et al. (2009), significantly younger than the age of 42.3±1.1 Ma given in the EIDB. Note that in all three cases, the differences between the original and revised ages are outside the uncertainties stated in the EIDB. Therefore, at least some of impact ages listed in the EIDB have significant systematic uncertainties. Here, we repeat the analysis of Rampino and Caldeira (2015), first on their list, then on what we call the "updated" list. We then also apply the CSA to a newly compiled list containing only impacts with robust ages, mostly from a compilation by (Jourdan, 2012; Jourdan et al., 2009), but also extended here with a few recently dated impacts with robust ages (Table 1; Fig. 2). Finally, we discuss the so far unrecognized role of what we call "clusters" (sets of impacts with overlapping ages) in a CSA.

## 2 CIRCULAR SPECTRAL ANALYSIS OF AN UPDATED LIST OF IMPACTS

First, we repeat the CSA from Rampino and Caldeira (2015) to see if we can retrieve the reported ~26 Ma peak. The CSA is done using a MATLAB script (available from the first author by request), following the descriptions given in Rampino and Caldeira (2015) and Lutz (1985). In contrast to Rampino and Caldeira (2015), we also account for overwrap, i.e., for the effect that the length of the time-line of impacts is in general not an integer multiple of the trial period $P_i$ (details given by Lutz, 1985). For the list given by Rampino and Caldeira (2015), we successfully retrieve a peak at ~26 Ma in the R versus P diagram (Fig. 3). The shape of the curve in the periodogram is slightly different in our analysis compared to the one presented in Fig. 1 of Rampino and Caldeira (2015), possibly due to our overwrap correction. Our SL (R =

**Table 1.** List of 26 impacts with robust ages. 17 were compiled by Jourdan et al. (2009); Jourdan (2012), and 9 additional, more recently dated impacts have been added here. All ages use the revised 40K decay constant given by Renne et al. (2010). Clusters: *65 Ma,**229 Ma,***462 Ma. Some impacts have multiple age references, see Jourdan (2012) for complete list.

| Impact structure | Diam. (km) | EIDB age (Ma), 1 SD | Robust age (Ma), 2 SE | Reference(s) |
| --- | --- | --- | --- | --- |
| Ries (+Steinheim) | 24 | 15.1±0.1 | 14.83±0.15 (1.01%) | (See Jourdan, 2012) |
| Logoisk | 17 | 42.3±1.1 | 30.0±0.5 (1.67%) | (Sherlock et al., 2009) |
| Chesapeake Bay | 40 | 35.3±0.1 | 35.67±0.28 (0.78%) | (See Jourdan, 2012) |
| Kamensk | 25 | 49.0±0.2 | 50.37±0.40 (0.79%) | (Izett et al., 1994) |
| Boltysh* | 24 | 65.17±0.64 | 65.82±0.74 (1.12%) | (Kelley and Gurov, 2002) |
| Chicxulub* | 170 | 64.98±0.05 | 66.07±0.37 (0.56%) | (Swisher et al., 1992) |
| Lappajärvi | 23 | 76.20±0.29 | 76.20±0.29 (0.38%) | (Schmieder and Jourdan, 2013) |
| Morokweng | 70 | 145.0±0.8 | 145.2±0.8 (0.55%) | (See Jourdan, 2012) |
| Puchezh-Katunki | 40 | 167±3 | 193.8±1.1 (0.57%) | (Holm-Alwmark et al., 2016) |
| Rochechouart | 23 | 201±2 | 202.7±2.2 (1.09%) | (Schmieder et al., 2010) |
| Manicouagan | 100 | 214±1 | 214.56±0.05 (0.02%) | (Jourdan, 2012) |
| Lake Saint Martin** | 40 | 220±32 | 227.8±1.1 (0.35%) | (Schmieder et al., 2014) |
| Paasselkä** | 10 | <1800 | 231.0±2.2 (0.78%) | (Schwarz et al., 2015) |
| Araguahina | 40 | 254.7±2.5 | 254.7±2.5 (0.98%) | (Tohver et al., 2012) |
| Clearwater West | 36 | 290±20 | 286.2±2.2 (0.77%) | (Schmieder et al., 2015) |
| Siljan | 52 | 376.8±1.7 | 380.9±4.6 (1.21%) | (Reimold et al., 2005) |
| Kaluga | 15 | 380±5 | 395±4 (1.01%) | (Masaitis, 2002) |
| Kärdla*** | 4 | ~455 | 461±5 (1.08%) | (See Jourdan, 2012) |
| Lockne (+Målingen)*** | 8 | ~458 | 461±5 (1.08%) | (See Jourdan, 2012) |
| Tvären*** | 2 | ~455 | 462±5 (1.08%) | (See Jourdan, 2012) |
| Clearwater East*** | 26 | 460-470 | 465±5 (1.08%) | (Schmieder et al., 2015) |
| Carswell | 39 | 115±10 | 485.5±1.5 (0.45%) | (Bleeker et al., 2015) |
| Jänisjärvi | 14 | 700±5 | 687±5 (0.73%) | (Jourdan et al., 2008) |
| Keurusselkä | 30 | <1800 | 1151±11 (0.87%) | (Schmieder et al., 2016) |
| Sudbury | 250 | 1850±3 | 1849.3±0.3 (0.02%) | (See Jourdan, 2012) |
| Vredefort | 300 | 2023±4 | 2023±4 (0.20%) | (See Jourdan, 2012) |

~0.28) is similar, but still above the one determined by Rampino and Caldeira (2015), likely because the random impact series they used to determine the SL have impact ages following a non-uniform (Γ-distributed) probability distribution. We prefer a uniform distribution because the list of impacts with reliable ages we use later on shows no clear trend in time (Fig. 2).

We then did a CSA on an "updated" version of the Rampino and Caldeira (2015) list, containing the following changes: (1) We removed Steinheim, which should not be counted as an independent event as it likely formed alongside the nearby (40 km) Ries impact structure when a binary asteroid impacted (Stöffler et al., 2002); (2) Carswell was removed as it has a revised age >260 Ma (Alwmark et al., submitted; Bleeker et al., 2015); (3) the age of Puchezh-Katunki was adjusted from 167 Ma to 193.8 Ma (Holm-Alwmark et al., submitted; Holm-Alwmark et al., 2016); (4) the age of Logoisk was also adjusted from 42.3 Ma to 30.0 Ma (Jourdan, 2012; Sherlock et al., 2009); (5) the Paasselkä and Lake Saint Martin impacts were added. The latter two impacts were not included by Rampino and Caldeira (2015) because the age uncertainties as given in the EIDB at the time when they compiled their list were larger than 10 Ma. However, both impact structures have recently been dated with high precision (Schmieder et al., 2014; Schwarz et al., 2015), and do now qualify for the list by the criteria given in Rampino and Caldeira (2015). The updated list thus contains again 37 impacts. As shown in Fig. 3, these changes result in a weakened ~26 Ma peak, falling slightly below our SL, although it is still above the SL (~0.23 at 26 Ma) given by Rampino and Caldeira, (2015). There are additional peaks above the SL at 7 and 11 Ma. Rampino and Caldeira (2015) dismiss them as being "below the Nyquist-limit", but the definition of this limit in a non-uniformly sampled record is not trivial (e.g., Maciejewski et al., 2009). We do not investigate this in more detail, but note that even if the impact distribution is random, on average 5\% of the R values will be above the SL (as it is defined by the 95$^{th}$ percentile). To illustrate this, we show the periodogram curve of an artificial impact series with uniform random age distribution (Fig. 3), which nevertheless shows similar sharp, short-period peaks. In the following, we will ignore the short-period (<20 Ma) peaks (as did Rampino and Caldeira, 2015).

## 3 CIRCULAR SPECTRAL ANALYSIS OF A LIST OF IMPACTS WITH ROBUST AGES

As shown in the last section, a CSA should not be applied to compilations of impact ages of variable quality, as even small changes (removing, adding and changing individual data points) can significantly change its outcome. An arguably better approach might be to work with a database of impacts with high quality ages, such as the one compiled by (Jourdan, 2012; Jourdan et al., 2009). In Table 1, we list the 17 impacts with robust ages from that database, and add 9 impacts with more recently published, robust ages: Araguainha (Tohver et al., 2012), Carswell (Alwmark et al., submitted; Bleeker et al., 2015), Clearwater West and Clearwater East (Schmieder et al., 2015), Keurusselkä (Schmieder et al., 2016), Lake Saint Martin (Schmieder et al., 2014), Lappajärvi (Schmieder and Jourdan, 2013), Paasselkä (Schwarz et al., 2015), and Puchezh-Katunki (Holm-Alwmark et al., submitted; (Holm-Alwmark et al., 2016)). The uncertainties of the individual ages in this list are significantly reduced compared to the list by Rampino and Caldeira (2015), with the average relative uncertainty a factor of 9 lower (0.8% vs. 7.2%, 2σ) and the maximum absolute uncertainty a factor of 2 lower (5 Ma) for impacts within the last 500 Ma. In the following, we will only consider either the 22 impacts with robust ages from the last 500 Ma (the period for which a periodicity of the impact/extinction event record has been discussed; e.g., Melott and Bambach (2010), or the 14 impacts with robust ages from the last 260 Ma (the period to which Rampino and Caldeira (2015) restricted their analysis). The results do not change qualitatively if the four Precambrian impacts are included in the analysis.

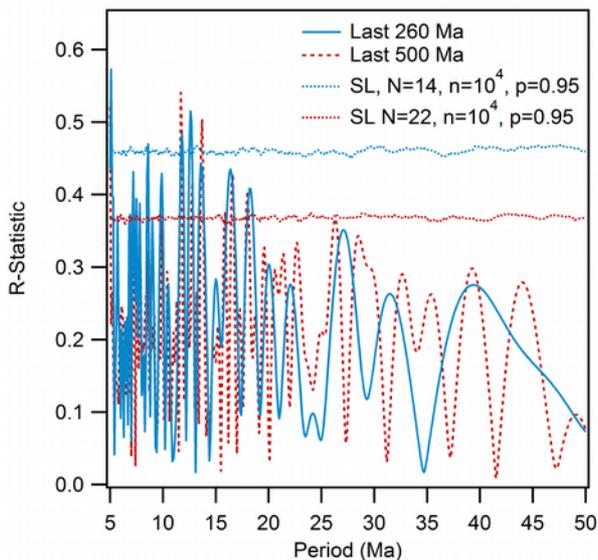

*Figure 4.* CSA for the list of impacts with robust ages given in Table 1 for the last 260 Ma (solid line), and the last 500 Ma (short-dashed line), respectively.

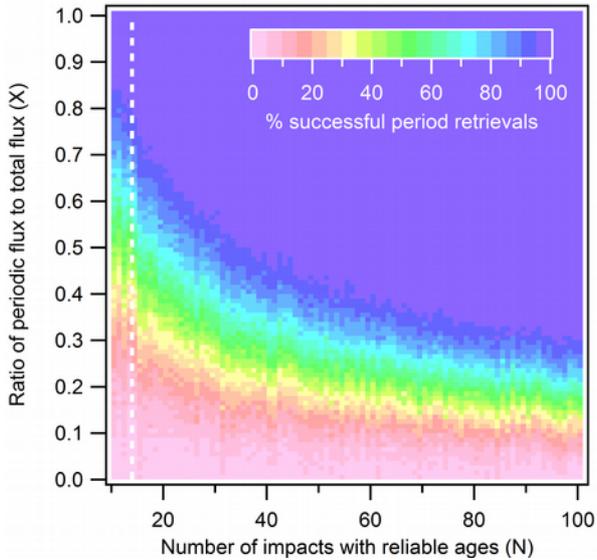

**Figure 5.** Percentage of runs (out of 500 trials for each data point) where a CSA resulted in the successful retrieval of an artificially induced 26 Ma period, for a given number of impacts (x-axis) and a given contribution of the periodic ux (y-axis). The vertical dashed line represents the 14 impacts with robust ages known from the last 260 Ma.

In Fig. 4, we show the result of the CSA of the list of robust impact ages in Table 1, for both the last 260 Ma, and the last 500 Ma. Due to the fewer impacts included in our list compared to the one of Rampino and Caldeira (2015) (14 for the 260 Ma interval, 22 for the 500 Ma interval, vs. 37 in the 260 Ma interval in Rampino and Caldeira (2015)), the SL is higher, at ca. 0.37, and 0.46, respectively. Again, we find no significant peak at ~26 Ma. It could now be argued that the reason for our failure to retrieve the peak found by Rampino and Caldeira (2015) is due to the fewer impacts considered. Furthermore, it could be argued that only a fraction of the impacts might have been produced by a periodic component, with the rest of the impacts distributed at random. To test both propositions, we extended our MATLAB script, to generate impact series composed of both an artificially induced periodic signal, and a random signal. A fraction X of the total impacts within a time interval (the last 5–260 Ma) were assigned randomly to the peaks of regular intervals (with a period P = 26 Ma, and a phase of 15 Ma, although the phase is of no consequence here), and then varied uniform-randomly within a relative age uncertainty range of 1%. The remaining impacts (1-X) were distributed uniform-randomly within the interval 5–260 Ma (i.e., a small fraction of those might add to the periodicity by pure coincidence). We then varied both the total number of impacts N (10–100; x-axis in Fig. 5) and the relative contribution X of the periodic fraction to the total flux (0–100%; y-axis in Fig. 5). For each combination of N and X, we created 500 "partially" periodic impact series, carried out a CSA on each series and checked whether the artificially induced periodicity was successfully recovered at $P_i$ = 26 Ma again (i.e., whether the R value at 26 Ma was above the 95$^{th}$ percentile of $10^4$ random runs with the same N, but X = 0%). The result of the simulation is shown in Fig. 5. The colours represent the fraction of runs in which the artificial periodicity was successfully recovered by the CSA (see legend). Clearly, 14 impacts are enough to recover a 26 Ma periodicity if X = 100%, i.e., if the impact record is fully periodic. For a scenario where the impact record is a combination of a periodic and a random component, the periodicity is still successfully recovered in 95% of the cases with N = 14 impacts as long as the periodic contribution X is higher than 80%. Therefore, if we consider CSA a useful method, not finding a periodicity within the set of 14 impacts with robust ages from the last 260 Ma (Fig. 4) allows us to exclude that more than 80% of the impacts are from a periodic component – otherwise, we should have found it by now.

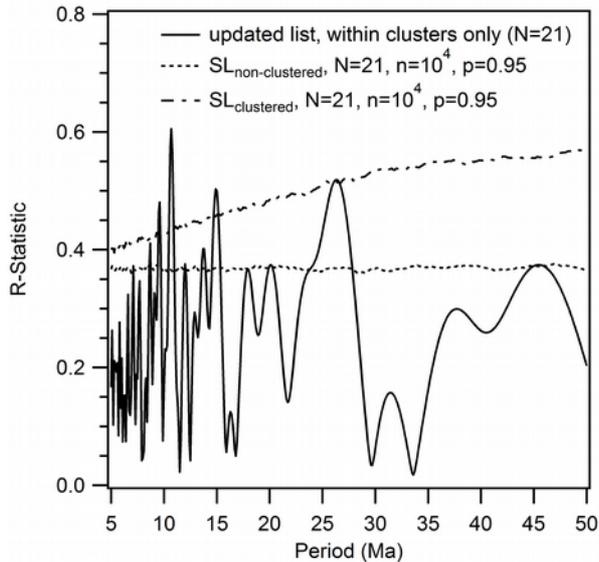

*Figure 6. CSA of only the clustered impacts in the list of Rampino & Caldeira (2015) updated by us. The dotted line represents the significance limit for an non-clustered series, whereas the dash-dotted line represents the significance limit for a clustered impact series.*

For N = 37 impacts, even an X = 50% should be recovered by a CSA in 95% of the cases. So it is possible, in principle, that a CSA on 37 impacts in the last 260 Ma could recover a periodicity in the impact record (as claimed by Rampino and Caldeira, 2015), while a CSA on 14 impacts of the same record (our analysis here) would not. In that case, X would likely need to be about 50 – 80% so that it is at the same time likely to be found with 37 impacts and unlikely to be found with only 14 impacts. As shown in Fig. 5, the determination of additional robust ages can eventually exclude (or confirm) an X between 50 – 80%: about 40 robust impact ages (within the last 260 Ma) are needed for X=50% (at a confidence level of 95%). Even in a fully random impact record of sufficient size, one can of course always select (or "cherry pick") a subset of impacts which fall into nearly periodic intervals, but this should never be considered "evidence" in favour of a small periodic component. What can be said with very high confidence today is that the impact record is not fully periodic (i.e., X ≠ 100%) and that if a periodic component exists, it must contribute less than <80% of the impact record, at a confidence level of 95%. However, the proposal of a lower value for X has to await the determination of additional robust impact ages before it can be seriously considered again.

## 4 CLUSTERED IMPACTS AND PERODICITY

Except for the two impacts added (Lake Saint Martin and Paasselkä), and the two impacts with adjusted ages (Logoisk and Puchezh-Katunki), our list of 14 impacts with robust ages from the last 260 Ma is essentially a subset of the 37 impacts in the original list of Rampino and Caldeira (2015). Therefore, the impacts with non-robust ages we excluded from our analysis are the likely "carriers" of the putative periodic signal reported by Rampino and Caldeira, (2015). Interestingly, many of those are found in what we call "clusters" (as did Jourdan, 2012). Within a cluster, all ages are mutually within the age uncertainties (2σ) of the other cluster members, i.e., all impacts within a cluster could be of the same age. Note that this definition of a cluster is not tied in any way to periodicity: both impacts in- and outside of clusters might contribute to a periodic signal (or not). In the Rampino and Caldeira (2015) list, such impact age clusters are found at ca. 36, 47, 65, 90, 120, 142 and 167 Ma. It is remarkable that with the exception of the one at 47 Ma, they are indeed spaced by about 22 – 30 Ma, roughly the range of the period suggested by Rampino and Caldeira (2015) and in previous works. Now, one might suspect that these clusters are simply due to chance and large age uncertainties. While this is certainly possible, impact age clusters are not restricted to non-robust ages: The list of robust impact ages in Table 1 also contains three clusters, one at the K-Pg boundary (Chicxulub and Boltysh, ca. 65 Ma; the only cluster present in both data sets), one of Late Triassic age (Lake Saint Martin and Paasselkä, ca. 229 Ma) and a large one of Middle Ordovician age (Kärdla, Lockne with Målingen, Tvären and Clearwater East, ca. 462 Ma). A potential fourth cluster could be found in the Late Eocene (ca. 35 Ma; Koeberl, 2009), but Chesapeake Bay is currently the only impact from that period with a robust age. We do not consider the Ries and Steinheim impacts to be in a cluster although their ages overlap, because, as mentioned above, they were formed by the impact of a binary asteroid (Stöffler et al., 2002), as were Lockne and Målingen (Ormö et al., 2014). Although the uncertainties in Table 1 are significantly smaller than the ones in the original list by Rampino and Caldeira (2015) (average relative 2σ uncertainties of 0.8% and 7.2%, respectively), we still find an important fraction of impacts in clusters (4 impacts out of 14 within the last 260 Ma, or 8 impacts out of 22 within the last 500 Ma). Based on a run of $10^5$ artificial impact series with uniform-randomly distributed ages, we estimate that the chance of finding 4 out of 14 impacts, or 8 out of 22 impacts within 0.8% uncertainty of each other (regardless over how many clusters they are distributed) is only 7.2%, and 5.6%, respectively.

It is thus possible that some of these clusters are real in the sense that they reflect a short but dramatic increase of the impact cratering rate on Earth. The most notable cluster of impacts with robust ages is found in the Ordovician (ca. 465±5 Ma ago), encompassing four of the impacts listed in Table 1, and many more candidates of similar (but non-robust) ages listed in the EIDB. A plausible explanation for this cluster is the break-up of a large asteroid near a powerful orbital resonance in the asteroid belt, resulting in a "shower" of fragments delivered to Earth-crossing orbits (Heck et al., 2004; Nesvorný et al., 2009; Ormö et al., 2014; Schmitz et al., 2001), confirmed independently by Ar-Ar ages of meteorites (Bogard, 2011), the recovery of fossil meteorites from Sweden (Heck et al., 2004; Schmitz et al., 2001) and micrometeoritic dust from different Ordovician sediment layers in Sweden, China and Russia (Alwmark et al., 2012; Meier et al., 2014, 2010). Nesvorný et al., (2009) suggested that the Gefion asteroid family in the main asteroid belt was formed by the same event. Numerous such asteroid break-up events have been identified for the last 260 Ma (e.g., Spoto et al., 2015), some of which may well have resulted in similar short-duration "impact showers" on Earth. Other possible causal connections between impacts with similar ages have been suggested, e.g. invoking comet showers or tidal break-ups of comets (similar to the collision of comet Shoemaker-Levy 9 with Jupiter), for the late Eocene (as reviewed by Koeberl, 2009), for the Boltysh and Chicxulub impacts at the Cretaceous-Paleocene boundary (Jolley et al., 2010), and for the Late Triassic (Schmieder et al., 2014; Spray et al., 1998).

If these clusters are indeed the result of occasional asteroid break-up events, we do not expect them to appear at periodic intervals. But a CSA done on just the 21 impacts within clusters (from the list of 37 impacts by Rampino and Caldeira (2015) updated by us here) results in a strong peak at ~26 Ma (Fig. 6). First, this confirms that it is indeed the subset of clustered impacts which "carries" the 26 Ma periodic signal identified by Rampino and Caldeira, (2015). But then, what is the meaning of this peak? Do we have to consider a possible periodicity in the formation of impact clusters (or asteroid break-up events)? If compared with the 95[th] percentile SL derived from 21 uniform randomly distributed impact ages ($SL_{non-clustered}$ in Fig. 6), the peak seems highly significant. But this is not the SL which should be used here. In order to test whether clusters (or rather the cluster age centres) appear at periodic intervals, the random distribution against which the observed distribution is tested has to be clustered as well, with randomly-spaced cluster age centres. We created a set of $10^4$ impact series with the centres of clusters at (uniform) random spacings, 2 – 5 impacts per cluster, a cluster size-frequency distribution scaled to replicate the one in the updated Rampino and Caldeira (2015) list (67% of the clusters containing 2 impacts, 67% of the remaining clusters having 3 impacts, etc.), the ages of impacts within each cluster varying by up to 4% from the cluster centre age, and each series containing 21 impacts in total. The 95th percentile of the clustered impact series population shows increasing R values with increasing period ($SL_{clustered}$ in Fig. 6). The shape of $SL_{clustered}$ depends on the scaling parameter used (here: 67%) and the age uncertainty allowed for cluster members (here: 4%), with lower values in both parameters leading to a steeper slope at shorter periods. If we use this clustered random model instead of the uniform random model to determine the SL, the 26 Ma peak arising among the 21 clustered impacts from the Rampino and Caldeira, (2015) list is no longer above the SL, i.e., the hypothesis of randomness can no longer be rejected at the 95% level (Fig. 6).

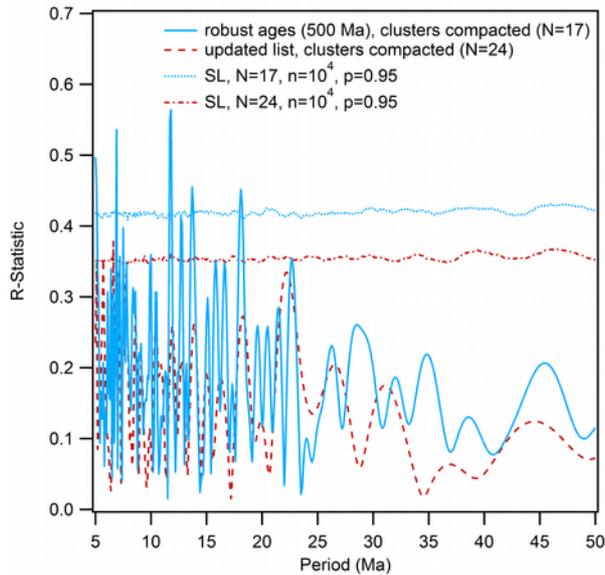

*Figure 7. CSA of the list of Rampino & Caldeira (2015) updated by us (dashed line) and the list of robust ages (this work; solid line). In both cases, all clustered impacts have been compacted into a single (mean) event age.*

Therefore, the presence of clusters in the impact record can lead to the erroneous identification of peaks in an R vs. P diagram. Since there is at least some evidence for the presence of clustered impacts in the terrestrial impact record, even among the impacts with robust ages

(e.g., in the Ordovician), the random model against which the periodic model is being tested in a CSA should also contain clusters. An alternative and arguably simpler approach is to compact all impacts within a cluster into a single data point (having the average age of all impacts within the cluster). If clusters do indeed have a common physical cause (e.g., an asteroid break-up event), this is a reasonable step, with the argument being essentially the same as the one used above to exclude doublet craters like Ries/Steinheim and Lockne/Målingen. In Fig. 7, we show the result of a CSA done on the updated Rampino and Caldeira (2015) list (containing 24 impacts after compaction of all clusters), and a CSA done on the list of robust ages (containing 17 impacts after compaction of all clusters). Neither of the two series shows a significant periodicity at 26 Ma.

## 5 CONCLUSIONS

As we have shown, caution is required when searching for periodicities in the impact record using the circular spectral analysis (CSA) method. Our CSA of an updated version of the list by Rampino and Caldeira (2015) already shows a much less significant peak at a period of ~26 Ma, after only a few changes motivated by recently revised impact ages. Furthermore, a CSA conducted on a list composed exclusively of robust impact ages (compiled by Jourdan, 2012; Jourdan et al., 2009) and extended here) over the last 260 and 500 Ma does not reveal any significant periodicity either. This can be used to exclude that more than 80% of the impacts in the last 260 Ma have been produced by a periodic component. Additional robust impact ages would be desirable to exclude (or confirm) the contribution of a periodic component at a lower level. The seemingly periodic signal identified by Rampino and Caldeira (2015) is contained within a subset of impacts with clustered ages. We have shown that the level of significance which has to be exceeded by an R-peak identified in a CSA has to be higher if the impact record contains clustered ages, and that the 26 Ma periodicity signal in the list of Rampino and Caldeira, (2015) is not significant at the 95th percentile level if this is taken into account. Therefore, there is currently no convincing evidence for the existence of a periodic component in the terrestrial impact record.

**ACKNOWLEDGEMENTS**

The authors thank T. Kenkmann and A. A. Plant for their comments and discussion, and C. Bailer-Jones for a thorough and helpful review. This work has been partially supported by a Swiss National Science Foundation Ambizione grant to MM.